\documentclass{aa}  

\usepackage{graphicx}   
\usepackage{txfonts}
\usepackage{natbib}
\bibpunct{(}{)}{;}{a}{}{,}
\usepackage{xspace}

\newcommand{\xmm}{{\it XMM-Newton}\xspace}

\newcommand{\ROSAT}{\hbox{ROSAT}\xspace}
\newcommand{\ero}{\hbox{eROSITA}\xspace}
\newcommand{\cxo}{\hbox{Chandra}\xspace}

\newcommand{\suzaku}{\hbox{Suzaku}\xspace}

\newcommand{\src}{{SN\,1987A}\xspace}

\newcommand{\ergcm}[1]{$\times 10^{#1}$ erg cm$^{-2}$ s$^{-1}$}
\newcommand{\oergcm}[1]{$10^{#1}$ erg cm$^{-2}$ s$^{-1}$}

\newcommand{\kms}{km s$^{-1}$}
\newcommand{\nh}{N$_{\rm H}$}

\newcommand{\ltsima}{$\buildrel < \over \sim$}
\newcommand{\lsim}{\lower.5ex\hbox{\ltsima}}
\newcommand{\gtsima}{$\buildrel > \over \sim$}
\newcommand{\gsim}{\lower.5ex\hbox{\gtsima}}

\begin{document} 

\title{SN\,1987A: Tracing the flux decline and spectral evolution through a comparison of SRG/\ero and \xmm observations}

\author{C.~Maitra\inst{1} \and
        F.~Haberl\inst{1} \and
        M.~Sasaki\inst{2} \and
        P.~Maggi\inst{3} \and
        K. Dennerl\inst{1} \and
        M.~J. Freyberg\inst{1} 
       } 

\titlerunning{SN\,1987A}
\authorrunning{Maitra et al.}

\institute{Max-Planck-Institut f{\"u}r extraterrestrische Physik, Gie{\ss}enbachstra{\ss}e 1, 85748 Garching, Germany, \email{cmaitra@mpe.mpg.de}
\and
 Remeis Observatory and ECAP, Universit{\"a}t Erlangen-N{\"u}rnberg, Sternwartstr. 7, 96049 Bamberg, Germany
 \and
 Universit\'e de Strasbourg, CNRS, Observatoire astronomique de Strasbourg, UMR 7550, F-67000 Strasbourg, France }

\date{Received ... / Accepted ...}

\abstract
   {\src is the supernova closest to us observed in the past four centuries. It provides the unique opportunity of witnessing the birth and evolution of a supernova remnant. Monitoring the source in X-rays  provides insights into the physics of the shock, the X-ray emitting plasma, and the interaction of the shock with the structures in the circumstellar medium. The source has been monitored by \xmm EPIC-pn from 2007--2020. SRG/\ero also observed the source during its commissioning phase and first light in September and October 2019.}   
   {We investigated the spectral and flux evolution of SN 1987A in X-rays in the last 14 years up to November 2020 using \xmm and \ero observations. }
   {We performed a detailed spectral analysis using a three-component plane-parallel shock model and analysed and modelled the EPIC-pn monitoring and \ero observations in a consistent manner. }
   {This paper reports a complete and the most recent flux evolution of \src in the soft (0.5-2\,keV) and hard (3-10\,keV) X-ray band. The flux in the soft band flattened around 9424\,d and then displayed a turnover between 10142--10493\,d, after which it showed a continued decline. At about the same time, a break in the hard-band flux time evolution slope was detected. This implies that the blast wave has now passed beyond the dense structures of the equatorial ring and is expanding farther into more tenuous circumstellar medium. The temporal evolution of the normalisations of the three shock components match the results of hydrodynamical simulations  well, which predict a blue supergiant progenitor scenario. The trend in recent epochs indicates that the emission caused by the forward shock after leaving the equatorial ring and by the reverse shock in the ejecta is now becoming more dominant. The elemental abundances in the hot plasma component are significantly higher than those in the cooler component, indicating an origin in the reverse shock that propagates into the ejecta.}
   {}

 
\maketitle   

\section{Introduction}
Located in the Large Magellanic Cloud (LMC) at a relatively close distance of 51.4$\pm$1.2\,kpc \citep{1999IAUS..190..549P}, supernova (SN) 1987A is the closest SN seen on Earth in the past four centuries. It provides the unprecedented opportunity of witnessing the birth and evolution of a SN into its remnant \citep[][and references therein]{2016ARA&A..54...19M}.  The study of \src is further intriguing because of the circumstellar ring system that surrounds it, which is composed of three co-axial rings. The rings either originated in the merging of a binary system about 20,000 yr before the explosion \citep{2007Sci...315.1103M} or in collisions of a fast wind of the blue supergiant phase of the progenitor with material from a slow wind of the earlier red supergiant phase \citep{2008A&A...488L..37C}. Thirty-four years after its explosion, \src continues to be intensely monitored at all wavelengths. Monitoring the source in X-rays is particularly interesting because it provides insights into the physics of the shock-emitting plasma and its interaction with the structures in the circumstellar medium (CSM). 

\src was first detected in soft X-rays with \ROSAT  \citep{1994A&A...281L..45B,1996A&A...312L...9H}.
Its X-ray emission has increased for about 25 years since the first detection.
A steep rise in soft X-rays was detected
after 2000 at $\sim$4500\,d after the explosion with the \cxo X-ray observatory \citep{2005ApJ...634L..73P}
and with \xmm \citep{2006A&A...460..811H}.
The increase in X-ray emission was obviously correlated with the encounter of the blast wave with the equatorial high-density ring created by the progenitor.
Studies following in the years thereafter using \suzaku and \xmm observations have confirmed the steep increase 
\citep{2009PASJ...61..895S,2012A&A...548L...3M}.
Until 2012 ($\sim$9000 d), the X-ray emission continued to rise. Soft X-ray emission below 2 keV increased with a higher rate than the harder emission (3 -- 10 keV).

The continuous monitoring has revealed that in the years around 2012, a change in the flux rise and in the X-ray spectrum occurred. This was also complemented by continuous optical monitoring observations, which indicated the passing of the blast wave over the dense equatorial structures and the formation of hotspots outside the equatorial ring around the same time \citep{2015ApJ...806L..19F,2019ApJ...886..147L}. In the X-ray regime, after the increase rate in flux started to slow down in 2006 ($\sim$7000 d), the X-ray light curve has significantly levelled off since 2012. A study of the
evolution of the flux and the morphology in X-rays over 16 years using \cxo was
presented by \cite{2016ApJ...829...40F}. 
While the soft emission below 0.8 keV increased until about 8000\,d and remained constant, with a possible indication of a decrease, the X-ray light curve above 0.5 keV continued to rise, which is in particular prominent above 2 keV. 
In the \cxo images,  a reversal of the east-west asymmetry in the X-ray emission is observed. The eastern part started to fade after $\sim$9500\,d.

Recently, new studies based on archival data of \xmm taken in the past $\sim$15 years have been published. They further confirmed the declining trend of the soft-emission component, while the harder component continued to increase \citep{2021arXiv210302612A,2021arXiv210303844S}. \cite{2021arXiv210302612A} reported an analysis of 
\xmm EPIC-pn and RGS data up to 2019 together with NuSTAR data, while the  
analysis of \xmm observations (both EPIC and RGS) was also reported by \cite{2021arXiv210303844S}.
 \cite{2021arXiv210303844S} modelled  the \xmm spectra assuming two thermal plasma emission components, one reproducing the emission from the shocked equatorial ring, and the other from the gas around the SN shocked by the blast wave. This model is similar to what was assumed by  \citet{2010A&A...515A...5S} and \citet{2016ApJ...829...40F}. The \xmm data, however, also show a significant excess at higher energies, which might indicate the existence of an additional high-temperature plasma component \citep{2010MNRAS.407.1157Z,2012ApJ...752..103D,2012A&A...548L...3M,2021arXiv210302612A}.

Spectra of \src have also been taken with grating spectrometers (\xmm RGS, \cxo LETG, HETG), which allow constraining the emission components and therefore shock models thanks to the high spectral resolution
\citep{2006ApJ...645..293Z,2009ApJ...692.1190Z,2010A&A...515A...5S,2012ApJ...752..103D}. In particular, with 
\cxo HETG monitoring observations, the evolution of the spectral components and the elemental abundances was studied in the past 10 years  (\cite{2020ApJ...899...21B,2021arXiv210302612A}).

To better understand the possible components of the X-ray emission and the evolution of the SN, a hydrodynamic simulation and modelling of the resulting X-ray spectrum was performed by \citet{2012ApJ...752..103D} based on plasma properties obtained from grating spectra taken with \cxo and \xmm. 
The latest hydrodynamic simulations were presented by \cite{2020A&A...636A..22O}, who compared different progenitor scenarios and predicted the evolution and dynamical and radiative properties of the remnant of \src. 

Recent observations have also been analysed in a search for indications of non-thermal emission in the \cxo + NuSTAR spectrum. \cite{2021ApJ...908L..45G} have analysed \cxo and NuSTAR data and concluded that the data might indicate the presence of a pulsar wind nebula. However, the possible non-thermal emission might also be emission from particles accelerated in the shocks. The non-thermal emission component, however, was not confirmed by \citet{2021arXiv210302612A} in the analysis of \xmm and NuSTAR data.

In this paper we report the recent results from the \xmm monitoring of \src until the end of 2020 and the SRG/\ero observations of the source taken during its commissioning phase and first light. \ero is very similar to \xmm in terms of their on-axis point spread functions (PSF) and the effective area in the 0.5--2.0\,keV range. \ero has a larger field of view (1 degree diameter compared to 0.5 degrees) without CCD gaps, which can be read out within only 50\,ms. This provides the possibility of using subpixel resolution, and the field of view is characterized by a temporally more constant particle background. For the analysis of SN1987A, the main advantage with respect to \xmm is the improved spectral resolution, especially with respect to the redistribution, which could be considerably suppressed and enables spectroscopic studies down to 150\,eV. Section~\ref{sec:observations} presents the observational data, Sect.~\ref{sec:spec} the spectral analysis of the \xmm and \ero observations, Sect.~\ref{sec:results} the results, and Sect.~\ref{sec:duscussion} and ~\ref{sec:conclusions} the discussion and conclusions.

\section{Observational data}
\label{sec:observations}
We have monitored \src regularly since 2007 with \xmm about once a year. The most recent observation were performed in November 2020.
An additional observation to allow cross-calibration studies between \xmm and \ero was performed in September 2019, six weeks before the first-light observation of \ero. Results from the monitoring and older archival \xmm observations of \src are presented in \citet{2006A&A...460..811H},\citet{2008ApJ...676..361H},\citet{2010A&A...515A...5S}, and \citet{2012A&A...548L...3M,2016A&A...585A.162M}. Maggi et al. 
followed the evolution of \src until December 2012. The details of all \xmm observations are summarized in Table~\ref{tabobsxmm}. Only observations with EPIC-pn in full-frame mode were used for this work. We did not use Obsid 0831810101 from 2019 November 27 because the observation was performed with EPIC-pn in large-window mode, which might introduce systematic flux differences.

\src was observed by \ero during the commissioning phase and during the first- light observational campaign \citep{2021A&A...647A...1P}. 
A summary of the observations can be found in Table~\ref{tabobsero}. During the commissioning observation, only the cameras of telescope modules (TMs) 3 and 4 were switched on and observed the sky, while during first light, all cameras were active.

\section{Data reduction and spectral analysis}
\label{sec:spec}
\subsection{\xmm}
\label{sec:xmmspec}
 \xmm/EPIC \citep[see][for pn and MOS,  respectively]{2001A&A...365L..18S,2001A&A...365L..27T} data were processed using the latest \xmm data analysis software SAS, version 18.0.0\footnote{Science Analysis Software (SAS): \url{http://xmm.esac.esa.int/sas/}}. The observations were inspected for high background flaring activity by extracting the high-energy light curves (7.0\,keV$<$E$<$15\,keV for EPIC-pn) with a bin size of 100\,s. Event extraction was performed with the SAS task \texttt{evselect} using the standard filtering flags (\texttt{\#XMMEA\_EP}) for EPIC-pn. We extracted the spectra from the EPIC-pn data of all observations since 2007 (Table~\ref{tabobsxmm}) in a consistent way (common to all observations) with the same extraction regions around source (circle with 30\arcsec\ radius) and background (nearby circle with same radius) and using single-pixel events (\texttt{PATTERN} = 0), which have the best energy resolution.

To obtain the most recent picture of the spectral evolution of \src, we followed the approach described in \citet[][and references therein]{2012A&A...548L...3M}.
 As spectral model we used the same three-component plane-parallel shock model as \citet{2012A&A...548L...3M}, which was originally used by \citet{2012ApJ...752..103D} for \cxo and \xmm spectra with a fixed-temperature warm component (kT=1.15 keV).

\begin{table*}
\caption{Details of the XMM-Newton EPIC-pn observations}
\label{table_EPICobs}
\centering
\begin{tabular}{lcccrr}
\hline\hline
\noalign{\smallskip}
ObsId & Obs. start date & Age \tablefootmark{a} & Exp. times \tablefootmark{b} & Flux (0.5--2 keV) & Flux (3--10 keV) \\
      &                 & (days) & (ks) & \multicolumn{2}{c}{$\left(10^{-13}\right.$ erg\,s$^{-1}$\,cm$\left.^{-2}\right)$} \\
\noalign{\smallskip}
\hline
\noalign{\smallskip}
0406840301 & 2007-01-17 &  7269 & 106.9/93.1 & 33.80$^{+0.87}_{-0.77}$ &  4.18$^{+0.27}_{-0.52}$ \\ \noalign{\smallskip}
0506220101 & 2008-01-11 &  7629 & 109.4/89.4 & 43.96$^{+1.01}_{-0.91}$ &  5.48$^{+0.31}_{-0.39}$ \\ \noalign{\smallskip}
0556350101 & 2009-01-30 &  8013 & 100.0/86.7 & 53.28$^{+1.06}_{-0.93}$ &  6.49$^{+0.35}_{-0.43}$ \\ \noalign{\smallskip}
0601200101 & 2009-12-12 &  8329 &  89.9/77.5 & 59.70$^{+0.69}_{-0.74}$ &  7.25$^{+0.40}_{-0.61}$ \\ \noalign{\smallskip}
0650420101 & 2010-12-12 &  8693 &  64.0/55.9 & 66.55$^{+0.89}_{-0.73}$ &  8.45$^{+0.47}_{-0.62}$ \\ \noalign{\smallskip}
0671080101 & 2011-12-02 &  9049 &  80.6/70.2 & 71.25$^{+0.70}_{-0.55}$ & 10.16$^{+0.49}_{-0.59}$ \\ \noalign{\smallskip}
0690510101 & 2012-12-11 &  9424 &  68.0/60.5 & 74.93$^{+0.64}_{-0.52}$ & 11.59$^{+0.57}_{-0.77}$ \\ \noalign{\smallskip}
0743790101 & 2014-11-29 & 10142 &  77.6/67.8 & 74.85$^{+0.50}_{-0.63}$ & 13.35$^{+0.55}_{-0.59}$ \\ \noalign{\smallskip}
0763620101 & 2015-11-15 & 10493 &  64.0/56.9 & 73.87$^{+0.66}_{-0.56}$ & 14.76$^{+0.66}_{-0.81}$ \\ \noalign{\smallskip}
0783250201 & 2016-11-02 & 10846 &  72.4/63.3 & 71.61$^{+0.57}_{-0.55}$ & 15.22$^{+0.61}_{-0.70}$ \\ \noalign{\smallskip}
0804980201 & 2017-10-15 & 11193 &  68.7/59.1 & 69.76$^{+0.49}_{-0.61}$ & 15.66$^{+0.55}_{-0.58}$ \\ \noalign{\smallskip}
0852980101 & 2019-09-06 & 11884 &  14.7/12.4 & 64.58$^{+1.32}_{-0.97}$ & 16.75$^{+1.27}_{-1.46}$ \\ \noalign{\smallskip}
0862920201 & 2020-11-24 & 12329 &  74.2/60.5 & 61.78$^{+0.67}_{-0.57}$ & 16.97$^{+0.50}_{-0.53}$ \\ \noalign{\smallskip}
\hline
\label{tabobsxmm}
\end{tabular}
\tablefoot{All observations were operated with the EPIC-pn camera in full-frame read-out
mode with the medium optical blocking filter. 
Fluxes are given with 90\% confidence range and are not corrected for absorption.
\tablefoottext{a}{Number of days since the explosion of SN\,1987A.}
\tablefoottext{b}{Total/filtered exposure times (after removal of high background intervals).}
}
\end{table*}
\begin{table*} 
\caption{\ero observations of \src} 
\begin{center}
\begin{tabular}{ccccccc} 
\hline\hline\noalign{\smallskip}
Obsid &      Observation time                      & \multicolumn{4}{c}{Net exposure (ks)} \\
       &                                      & TM1  & TM2  & TM3  & TM4  & TM6 \\ 
\hline\noalign{\smallskip}
700016 (\ero Comm) & 2019-09-15 09:28 $-$ 2019-09-16 21:30 & --   & --   & 101  &  98  & --   \\ 
700161  (\ero FL) & 2019-10-18 16:54 $-$ 2019-10-19 15:08 & 71.1 & 71.4 & 71.4 & 71.4 & 71.4 \\ 
\hline
\label{tabobsero} 
\end{tabular} 
\end{center}
\end{table*}

The spectra were fitted simultaneously in the energy range of 0.2--10\,keV with the three-temperature shock model (\texttt{vpshock} with NEI version 3.0.4 in XSPEC version 12.11.0k). The temperatures of the cooler and hot components were allowed to vary {between epochs}, while parameters such as elemental abundances (N, O, Ne, Mg, Si, S, and Fe) were fitted in common to the spectra and tied between the different shock components. The redshift was fixed to a value of 285.4\,\kms\ , as in the previous works. The normalisations of all shock components were free for all spectra. Two more free parameters were removed by constraining the ionisation ages to cover a factor of 2 ($\tau_{lo}=\sqrt{2}\tau_{mid}$ ; $\tau_{high}=\tau_{mid}/\sqrt{2}$) similar to \citet{2012ApJ...752..103D}. In order to account for the photo-electric absorption, we included two XSPEC model components \texttt{phabs} and \texttt{vphabs}. The first, \nh$_{Gal}$ , was fixed to the value of $6\times10^{20}$\,cm$^{-2}$ \citep[][corresponding to Galactic foreground absorption]{1990ARA&A..28..215D}, and a second component was added to take the absorption inside the LMC into account. In the case of the second component, the metal abundances were fixed to 0.5 of the solar value \citep[][average metallicity in the LMC]{1992A&A...262...97V}. The \nh\ was tied between the different spectra.
We also included a Gaussian component to fit the Fe K$\alpha$ line in the spectra, with the line centroid energies and normalisations left free. The Fe K complex, itself a sum of several ionisation states of iron, is well reproduced by essentially only the hot component (see Fig.~\ref{fig_spec_pn_cmp} with individual components shown).
The origin of the Fe line complex in \src and its possible implications for the Fe abundance of the shocked plasma have been discussed previously in \cite{2010A&A...515A...5S}, \cite{2012A&A...548L...3M}, and \cite{2021arXiv210302612A}. However, we do not report these results in this work and leave it for a future detailed analysis. The errors are quoted at 90\% confidence throughout the paper. The upper limits are quoted at $3\sigma$ confidence. 

\subsection{\ero}
The data were reduced with a  pipeline based on the \ero Standard Analysis Software System (eSASS, Brunner et al., 2021, in preparation). We used data of the processing configuration c001,
which determines good time intervals, corrupted events and frames, and dead times, masks bad pixels, and applies pattern recognition and energy calibration. Finally, star-tracker and gyro data were used to assign celestial coordinates to each reconstructed X-ray photon. Circular regions with a radius of 75\arcsec\ and 100\arcsec\ were used for the source and background extraction, respectively. We rebinned the spectra to a minimum of 20 counts per bin in order to allow the use of the $\chi^{2}$-statistic. 
We extracted \ero spectra from \src from the available TMs from the calibrated event files using \texttt{srctool} with standard parameters suitable for point-source extraction. The background region was selected to be consistent with that chosen for the spectral analysis of the \xmm EPIC-pn spectra obtained from our monitoring program since 2007 (see Table~\ref{tabobsxmm}).
For EPIC-pn, we selected single-pixel events (\texttt{PAT} = 1), which provide
the best energy resolution. However, given the smaller pixel size of the \ero
CCDs, the fraction of single-pixel events is reduced, and we therefore also
created spectra using all valid pixel-patterns (\texttt{PAT} = 15) to use maximum statistics.
For the spectral analysis, we used only TM1, 2, 3, 4, and 6, which are equipped with cameras with on-chip optical blocking filters.

\section{Results}
\label{sec:results}
\subsection{Simultaneous fitting of EPIC-pn spectra from \xmm monitoring observations}

\begin{figure*}
  \begin{center}
   \resizebox{0.8\hsize}{!}{\includegraphics[angle=-90]{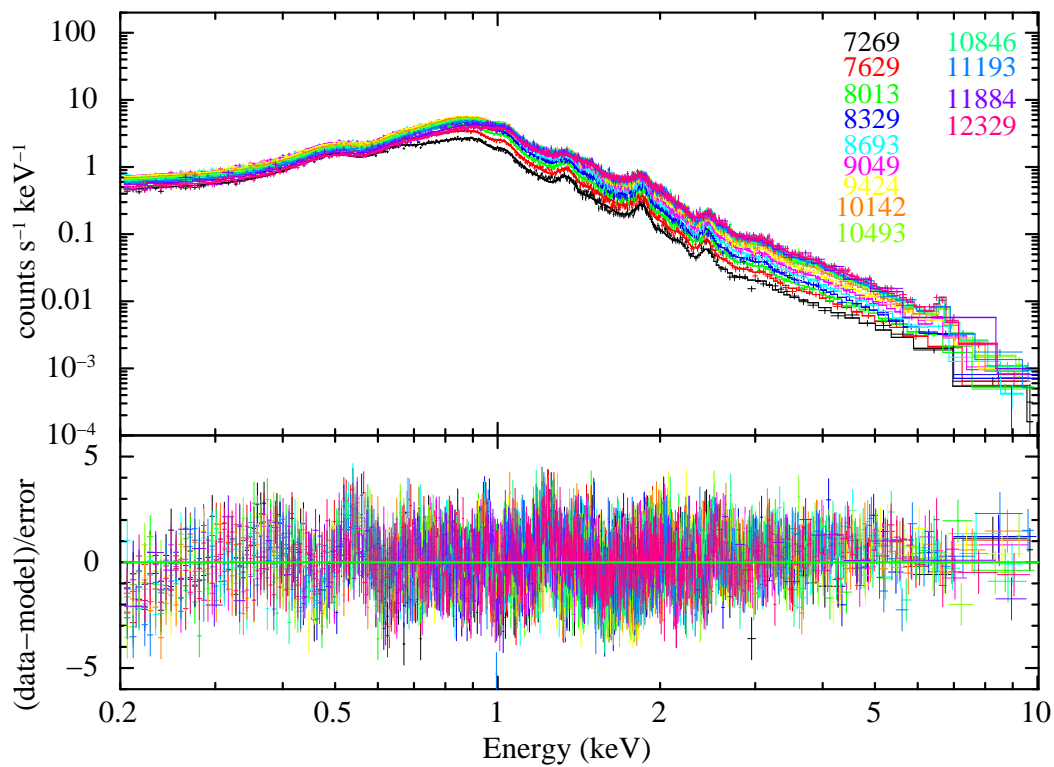}}
  \end{center}
  \caption{
    EPIC-pn spectra from thirteen monitoring observations of \src since 2007. The observation epochs (in days after the explosion) are marked in the figure, see Table.~\ref{tabobsero} for details. The spectra have been rebinned for visual clarity.}
  \label{figpn}
\end{figure*}

\begin{figure*}
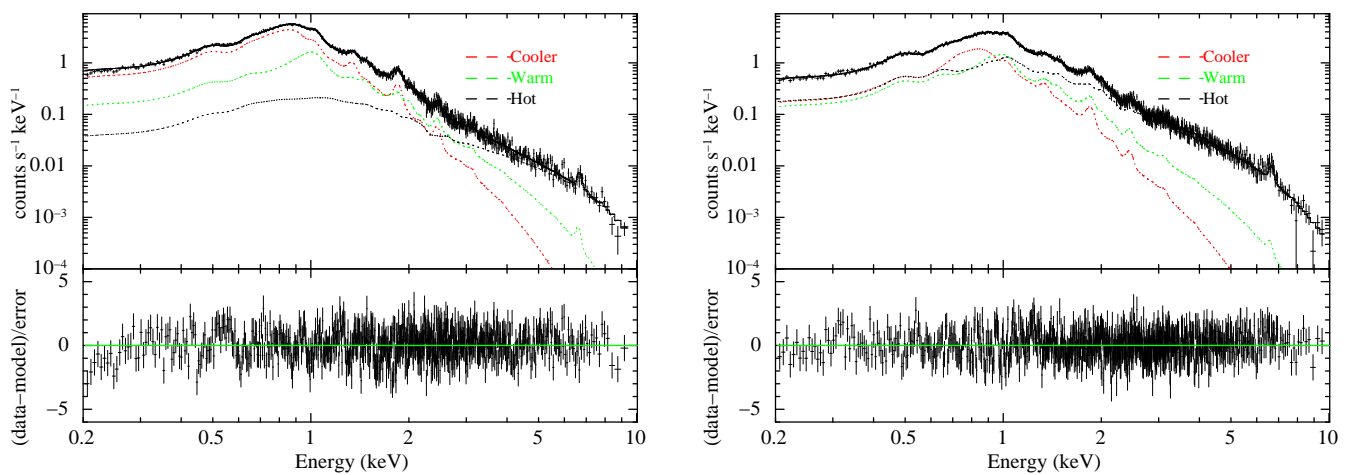

  \begin{center}
  \resizebox{0.49\hsize}{!}{\includegraphics[angle=-90]{spec_1987A_2012.eps}}
  \resizebox{0.49\hsize}{!}{\includegraphics[angle=-90]{1987A_2020.eps}}
 \end{center}
 \caption{EPIC-pn spectra from day 9424, Obsid 0690510101 (left) and day 12329, Obsid 0862920201 (right), showing the different model components: cooler (red), warm (green), and hot (black). }
  \label{fig_spec_pn_cmp}
\end{figure*}

EPIC-pn spectra with their best-fit models from the 13 monitoring observations of \src from 2007--2020 are shown in Fig.~\ref{figpn}. Figure~\ref{fig_spec_pn_cmp} shows the comparison between the EPIC-pn spectra from two different epochs. In 2012, epoch 9424\,d (left; where the cooler plasma component dominates), and the most recent epoch from 2020, 12329\,d (right; where the warm and hot plasma components begin to dominate the cooler component). The two epochs demonstrate that the flux and spectrum of \src have evolved significantly in the past few years.
The observed fluxes in the standard 0.5-2\,keV and 3-10\,keV band for each epoch are tabulated in Table~\ref{tabobsxmm}.
The elemental abundances of \src\, averaged over the epochs are provided in Table~\ref{tab:abundances}. Spectral parameters of the three-temperature shock model from EPIC-pn epochs 2007 and 2019 are listed in Table~\ref{tabspec}. The fitted elemental abundances for \src are slightly higher than obtained from \cxo HETG/LETG observations \citep[e.g.][]{2009ApJ...692.1190Z,2021arXiv210300244B}. The obtained range broadly agrees with the reported values in \citet{2010ApJ...717.1140M},\citet{2010A&A...515A...5S}, and \citet{2012ApJ...752..103D}, however.
This is further discussed in Sect.~\ref{sec:duscussion}. 

\begin{figure*}
  \begin{center}
  \resizebox{0.49\hsize}{!}{\includegraphics[angle=-90]{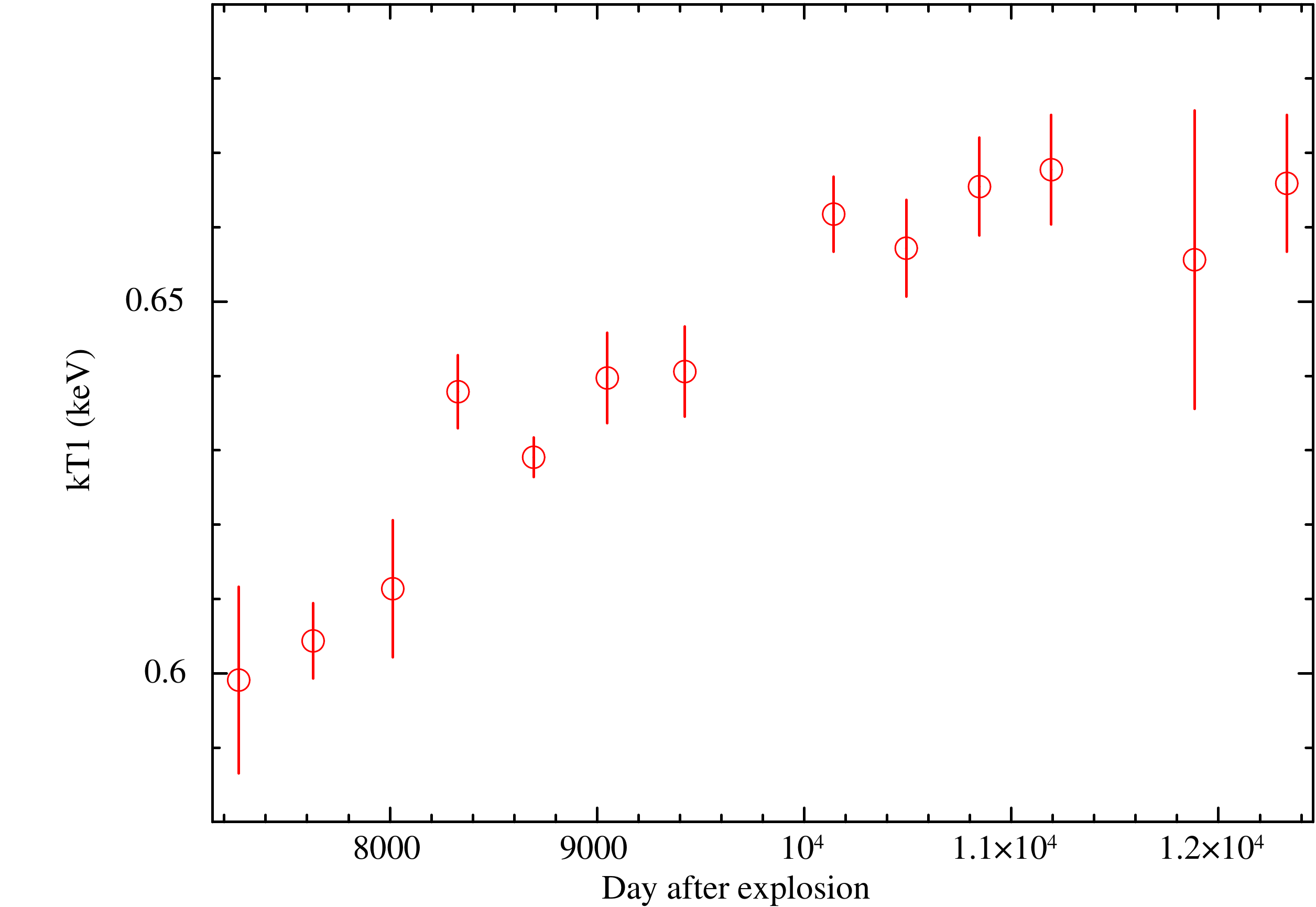}}
  \resizebox{0.49\hsize}{!}{\includegraphics[angle=-90]{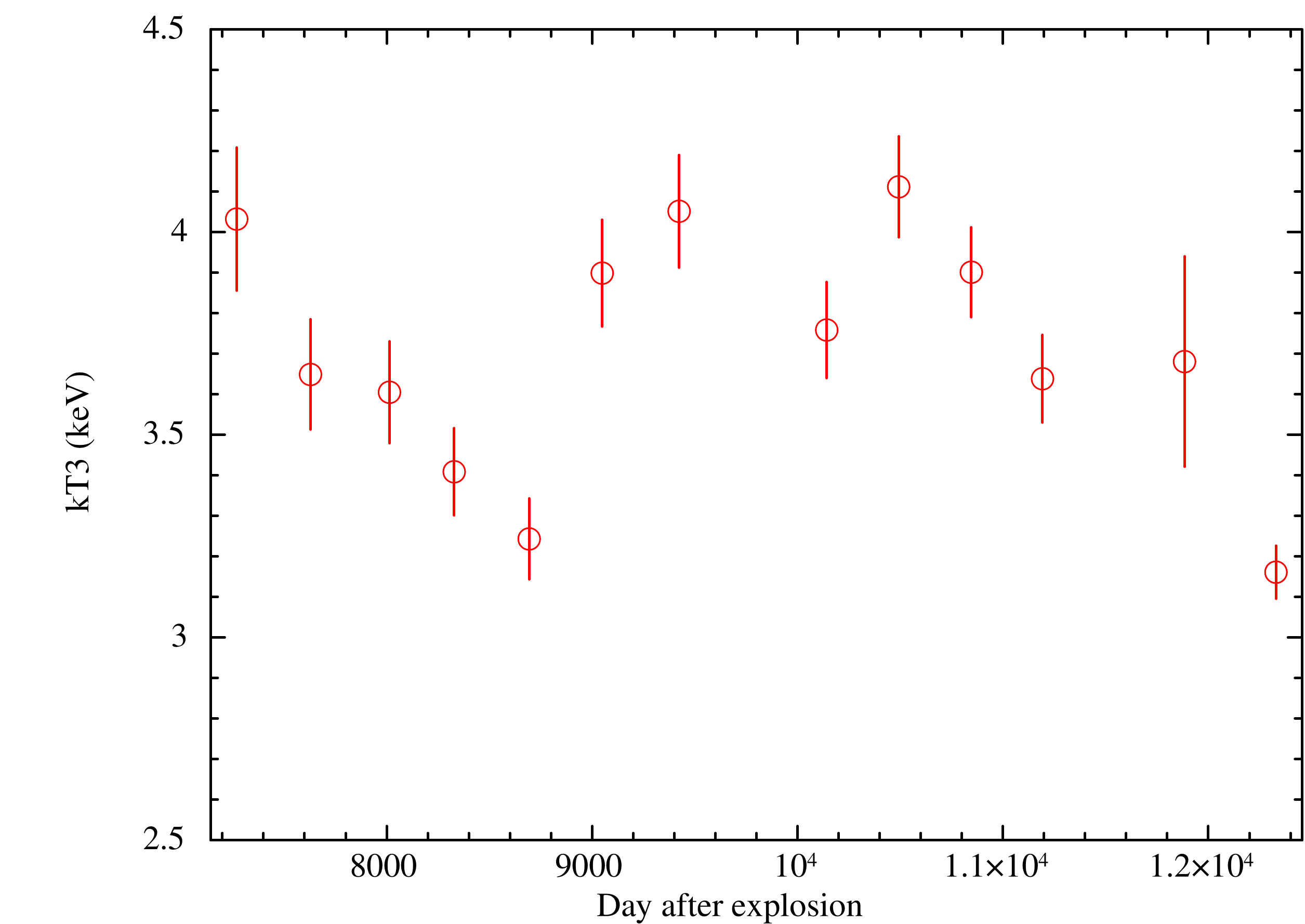}}
  \end{center}
  \caption{Evolution of the post-shock temperatures $kT_1$ of the cooler (left) and $kT_3$ hot (right) plasma components obtained from the EPIC-pn monitoring of \src.}
  \label{fig1987Atemp}
\end{figure*}

\begin{figure*}
  \begin{center}
  \resizebox{0.50\hsize}{!}{\includegraphics[angle=-90]{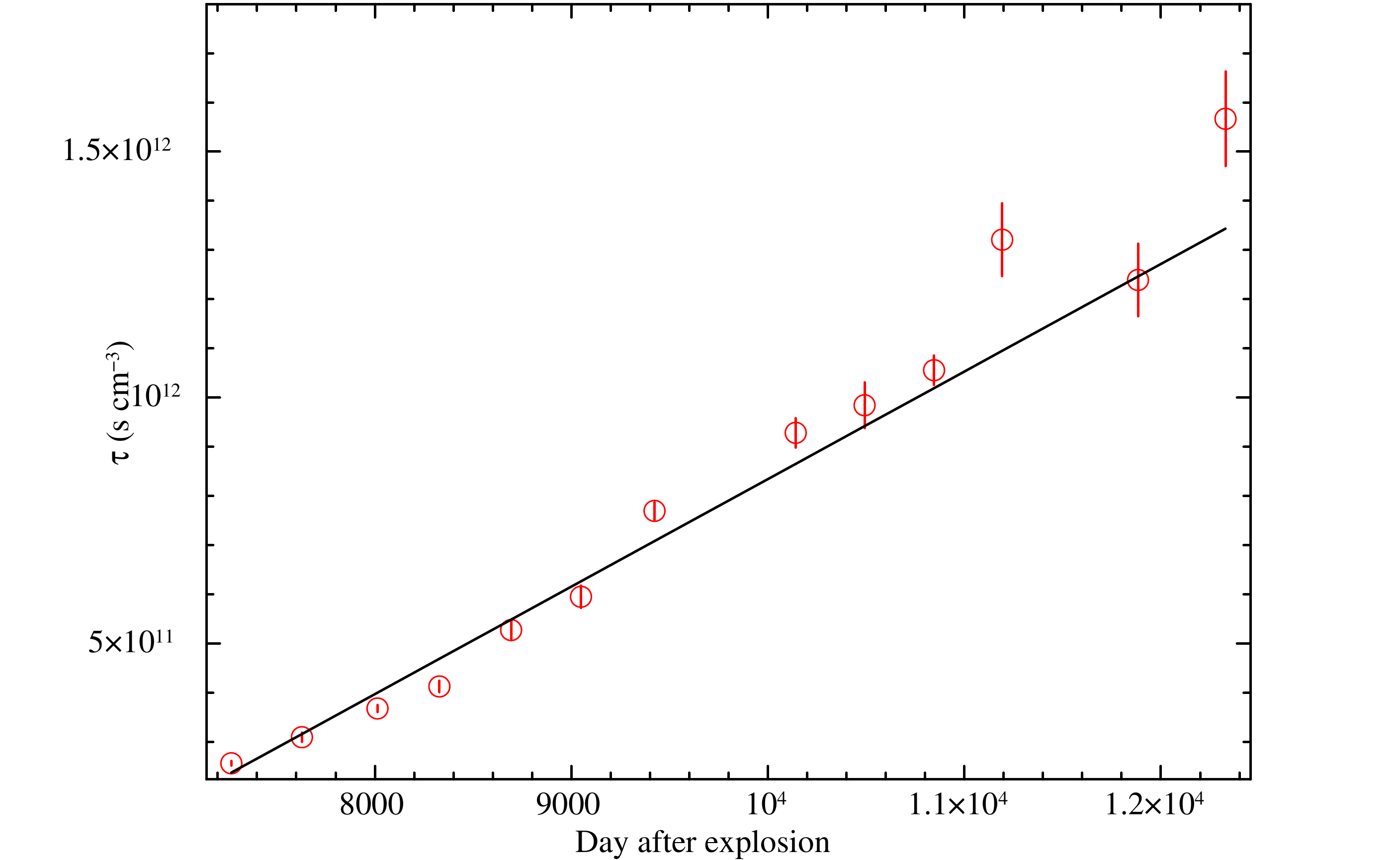}}
  \resizebox{0.49\hsize}{!}{\includegraphics[angle=-90]{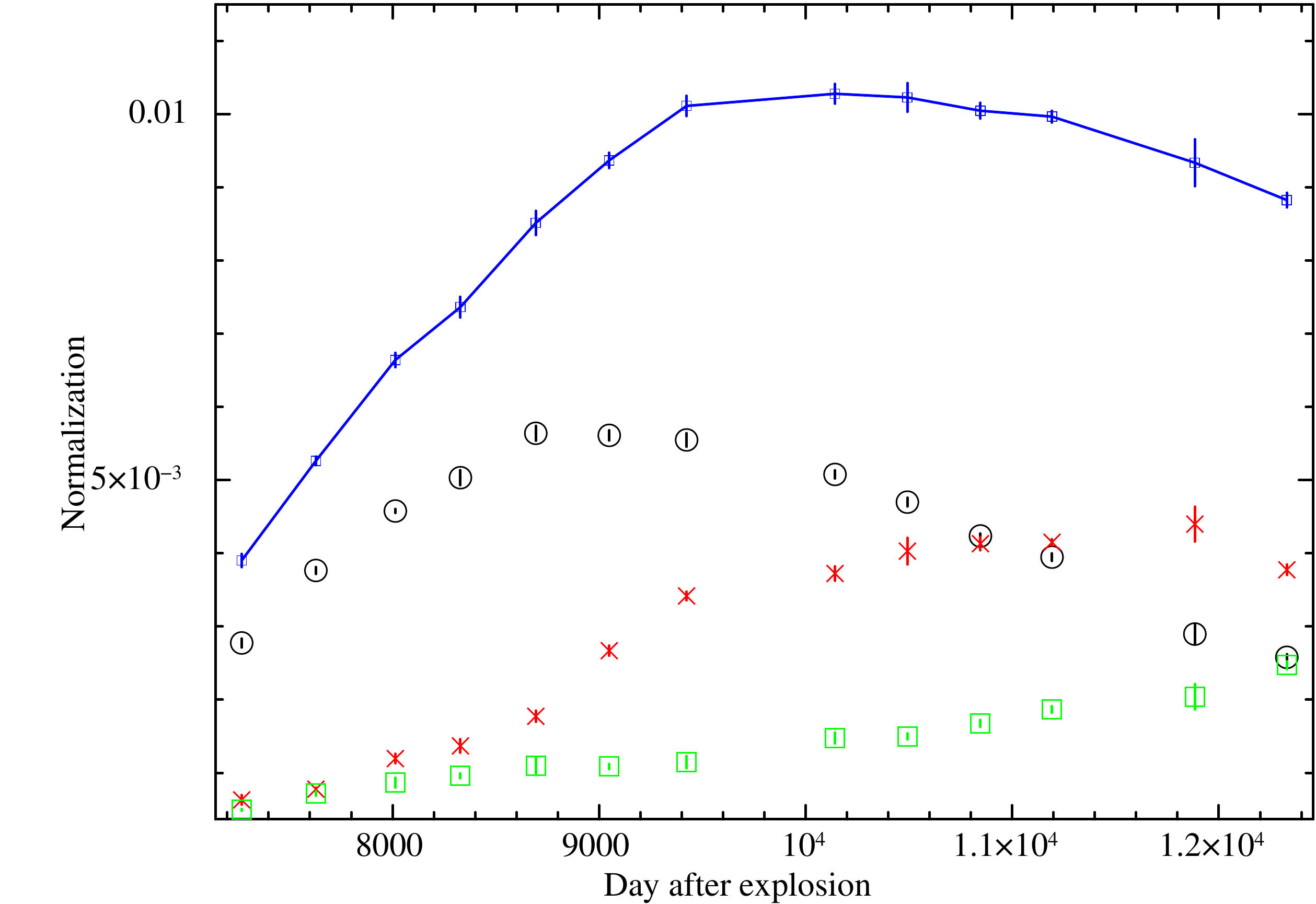}}
  \end{center}
  \caption{Evolution of the upper limit of the ionisation timescale $\tau_{{\rm u,}2}$ of the warm plasma component (left) and the \texttt{VPSHOCK} normalisations of \src obtained from the EPIC-pn monitoring (right). Black, red, and green symbols show the norm1, norm2, and norm3 components denoting the emission measures as defined in Table~\ref{tabspec}. The blue line marks the sum of the three components.}
  \label{fig1987Anorms}
\end{figure*}

We also investigated the possible temporal evolution of the post-shock temperatures of the cooler and hot plasma components, their normalisations and the upper limit of the ionisation timescales of the warm component (left free in the fit, as described in Sect.~\ref{sec:xmmspec}), as shown in Figs.~\ref{fig1987Atemp} and \ref{fig1987Anorms}.

Furthermore, in order to investigate possible differences in the elemental abundances between the three plasma components or their evolution with time, we performed several tests. 
First, we decoupled the abundances of the  cooler, warm, and hot plasma components separately (to avoid too many free parameters that would ensue if we were to decouple all three components at the same time).  Setting the cooler, warm, and hot plasma components free resulted in improvements of $\Delta\chi^{2}$=242, 130 and 384 for 6 degrees of freedom, respectively, indicating that the chemical composition of the hot component is most significantly different from the cooler and warm plasma components. The abundance of N could not be constrained for the warm and hot plasma components and was coupled in all cases. The fitted elemental abundances in all three cases are listed in Table~\ref{tab:abundances_2} and clearly show a trend of higher abundances for the warm and and especially the hot gas component. Furthermore, as the relative contribution of the three plasma components with respect to the total emission evolves with time between the epochs  (Fig.~\ref{fig1987Anorms}), we also fitted the elemental abundances in two groups corresponding to group 1 from 2007--2015, when the shocked plasma from the dense equatorial ring dominated, and group 2 from 2016--2020, when the warm and hot plasma components began to dominate. The results are compared to the average abundances in Table~\ref{tab:abundances}. The elemental abundance of N is significantly lower in the latest epoch, when the abundances of Ne and Mg increase marginally. No clear inference on the other elements can be drawn from this exercise.

Because of the long exposure that is available with EPIC-pn monitoring observations of \src for over a decade, we further tested whether a two-component shock model could be statistically ruled out in favour of the three-component model used in this work. 
For this purpose, we fitted all the EPIC-pn spectra simultaneously with only two \texttt{VPSHOCK} components. The ionisation timescales were left free for both, only the abundances were tied between the two components. The two-component fit results in a $\chi^{2}$ of 11860.9, to be compared with 11211.6 in case of the three-component model (used in this work). Both models have the same number of 9346 degrees of freedom due to the tied parameters in the latter case. This indicates that the three-component plane-parallel shock model used in this work is favoured by the data and further supports the findings of \citet{2012ApJ...752..103D}, who originally proposed the three-temperature shock model as they found a slightly better fit over a two-temperature model using the \xmm RGS and \cxo HETG grating instruments. Multi-temperature component models merely approximate a continuous distribution of temperatures in the plasma \citep{2015ApJ...810..168O}. However, it is clear that our X-ray data require the inclusion of this third hot component.

We also searched for an additional non-thermal emission from either a putative pulsar wind nebula or from synchrotron radiation \citep{2021ApJ...908L..45G}. To do this, we added a power-law component with index fixed at 2.0. The addition of the component does not lead to an improvement in the delta $\chi^{2}$ , indicating that the addition of this model is not statistically significant. The $3\sigma$ upper limit on the 3.0-10.0\,keV flux contributed by the power-law component is 1.9\ergcm{-15}. This is three orders of magnitudes lower than the flux obtained by \citet{2021ApJ...908L..45G} for a possible pulsar wind nebula ($F$(1-10\,keV) = \oergcm{-12}), assuming a model consisting of only two thermal-emission components and a power-law component. 
However, it should be noted that \cite{2021ApJ...908L..45G} cited significant absorption of this component in the surrounding cold ejecta, with an equivalent H column density of $\gtrsim10^{23}$\,cm$^{-2}$ , and the existence of a non-thermal emission cannot be conclusively ruled out using \xmm alone.
Similarly, addition of a very hot thermal component with $kT=20$\,keV is not favoured by the spectral fit, and the upper limit to the 3.0-10.0\,keV flux is 9\ergcm{-14}.


\begin{table}
\caption[]{Elemental abundances of \src from \xmm in various epochs. {The epoch ranges are mentioned in the table caption.} The elemental abundances are tied between different shock components.}
\begin{center}
\begin{tabular}{lccc}
\hline\hline\noalign{\smallskip}
\multicolumn{1}{l}{Element} &
\multicolumn{1}{c}{EPIC-pn} &
\multicolumn{1}{c}{EPIC-pn} &
\multicolumn{1}{c}{EPIC-pn} \\
\multicolumn{1}{l}{ } &
\multicolumn{1}{c}{2007--2020} &
\multicolumn{1}{c}{2007--2015} &
\multicolumn{1}{c}{2016--2020} \\
\noalign{\smallskip}\hline\noalign{\smallskip} 
N  &    4.23$\pm$0.12  & 4.13$\pm0.15$ & 2.77$_{-0.07}^{+0.64}$ \vspace{1mm} \\
O  &    0.33$\pm$0.01  & 0.32$\pm0.01$ & 0.27$\pm0.02$ \vspace{1mm}\\   
Ne &    0.66$_{-0.05}^{+0.01}$ & 0.64$_{-0.02}^{+0.03}$ \vspace{1mm} & 0.83$_{-0.06}^{+0.02}$ \\      
Mg &    0.64$\pm$0.01  & 0.63$\pm0.01$ & 0.71$_{-0.01}^{+0.02}$ \vspace{1mm}\\  
Si &    0.82$\pm$0.01  & 0.82$_{-0.02}^{+0.01}$ & 0.82$\pm0.02$ \vspace{1mm}\\  
S  &    0.65$\pm$0.02  & 0.67$_{-0.02}^{+0.03}$ & 0.57$\pm0.04$ \vspace{1mm}\\  
Fe &    0.38$\pm$0.01 & 0.38$\pm0.01$ & 0.39$\pm0.02$ \vspace{1mm}\\    
\hline
\end{tabular}
\end{center}
\label{tab:abundances}
\end{table}

\begin{table*}
\caption{Best-fit spectral parameters of \src for 2007 EPIC-pn, 2019 EPIC-pn, \ero Comm, \ero FL, \ero Comm+FL, and combined \ero +EPIC-pn 2019 observation with a three-component \texttt{VPSHOCK} model.}
\begin{center}
\begin{tabular}{lcccccc}
   \hline\hline\noalign{\smallskip}
   Parameter & EPIC-pn & EPIC-pn &  \ero & \ero & \ero Comm & \ero \\
        & 2007 & 2019 & Comm &  FL & + FL & + EPIC-pn 2019 \\
   \noalign{\smallskip}\hline\noalign{\smallskip}
   $N_H ^{{\rm LMC}}$ (10$^{22}$ cm$^{-2}$) & 0.244$\pm$0.002 & 0.244$\pm$0.002 &  0.26$\pm$0.02   & 0.24~$\pm0.01$ &  0.25~$\pm0.01$ & 0.25~$\pm0.01$ \\
   $kT_1$ (keV)      &   0.59$\pm$0.01                     & 0.66$\pm$0.02  & 0.67$\pm$0.03 &  0.69$\pm0.01$ & 0.68$\pm0.01$ & 0.68$\pm0.01$ \\
   $\tau_{{\rm u,}2}^{(b)}$     & 0.259$_{-0.002}^{+0.002}$            & 1.21$_{-0.05}^{+0.10}$ & 0.90$_{-0.23}^{+0.36}$ & 2.5$\pm0.5$ & 1.79$\pm0.27$ & 1.77$\pm0.26$ \\
   $kT_2$ (keV)  (frozen)   & 1.15                 & 1.15     & 1.15      & 1.15 & 1.15 & 1.15 \\
   $kT_3$ (keV)       &   4.04$\pm$0.26                     & 3.68$\pm$0.18 & 2.41$_{-0.11}^{+0.12}$  & 3.22$_{-0.20}^{+0.07}$ & 3.06$_{-0.12}^{+0.17}$ & 3.05$_{-0.12}^{+0.11}$ \\
   $norm_{1}^{(a)}$      &  2.7$\pm$0.1         & 2.9$\pm$0.1            & 3.5$_{-0.3}^{+0.2}$ &  3.1$_{-0.000.1}
  ^{+0.2}$ & 3.2$_{-0.4}^{+0.5}$ & 3.2$_{-0.1}^{+0.1}$ \\
   $norm_{2}^{(a)}$     &   0.63$_{-0.01}^{+0.04}$           & 3.8$_{-0.1}^{+0.4}$ & 1.3$_{-0.1}^{+0.1}$ & 3.4$_{-0.3}^{+0.4}$     & 3.4$_{-0.3}^{+0.3}$ &  3.4$_{-0.3}^{+0.3}$  \\
   $norm_{3}^{(a)}$     &   0.49$_{-0.01}^{+0.02}$          & 2.0$_{-0.1}^{+0.2}$ & 4.3$_{-0.6}^{+0.5}$ & 2.7$_{-0.3}^{+0.3}$ & 2.9$_{-0.2}^{+0.2}$ & 2.9$_{-0.2}^{+0.2}$ \\
  $\chi^{2}$    & 882.05 &  547.5  & 725.04  & 3004.7 & 2417.6 & 4189.4\\
  d.o.f. & 716 &511  & 571 & 2463 & 1941 & 3185 \\
   \noalign{\smallskip}\hline\noalign{\smallskip}
   \label{tabxayspec}
\end{tabular}
\tablefoot{
  Errors for $\Delta\chi^2=2.71$ confidence range.\\
  a: The normalisation is defined as (10$^{-14}$/4$\pi$$d$$^2$)$\int$$n$$_{e}$$n$$_{p}$$d$$V$, where $d$ is the distance to the SNR (in units of cm), $n$$_{\rm{e}}$ and $n$$_{\rm{p}}$ are the number densities of electrons and protons, respectively. Expressed on a scale of $10^{-3}$.\\
  $\tau_{\rm u}$: Upper limit on the ionisation time range in $10^{12}$ cm$^{-3}$ s. $\tau_{1}$ and $\tau_{3}$ are linked as $\tau_{2}=\frac{\tau_{1}}{\sqrt{2}}$ \& $\tau_{3}=\frac{\tau_{2}}{\sqrt{2}}$ \citep{2009ApJ...692.1190Z,2012ApJ...752..103D}
and thus we do not report \texttt{vpshock} parameters for it. 
}
\end{center}
\label{tabspec}
\end{table*}

\begin{table*}
\caption[]{Elemental abundances of \src from \xmm combining all epochs:  The first and second column represent the abundances obtained by untying the warm plasma component from the cooler and hot components, which are tied to teach other. The third and fourth columns represent the abundances obtained by untying the hot plasma component from the cooler and warm components, which are tied to teach other.  The fifth and the sixth column represent the same obtained by untying the cooler plasma component from the warm and hot components, which are tied to each other.}
\begin{center}
\begin{tabular}{lcc|cc| cc}
\hline\hline\noalign{\smallskip}
\multicolumn{1}{l}{Element} &
\multicolumn{1}{c}{cooler + hot plasma} &
\multicolumn{1}{c}{warm plasma} &
\multicolumn{1}{c}{cooler + warm plasma} &
\multicolumn{1}{c}{hot plasma} &
\multicolumn{1}{c}{warm + hot plasma} &
\multicolumn{1}{c}{cooler plasma} \\
\noalign{\smallskip}\hline\noalign{\smallskip} 
O  &    0.17$\pm$0.01   & 0.56$_{-0.03}^{+0.05}$ & 0.24$\pm0.02$       & 0.46$\pm0.04$ & 0.35 $\pm$ 0.01 & 0.22$\pm$0.01   \vspace{1mm}\\        
Ne &    0.24$\pm0.03$ & 2.15$_{-0.06}^{+0.46}$ & 0.39$\pm0.04$ & 3.16$\pm0.05$ &  1.74$_{-0.08}^{+0.12}$ &  0.24$_{-0.02}^{+0.03}$ \vspace{1mm}\\      
Mg &    0.42$\pm$0.03 & 1.18$_{-0.06}^{+0.11}$  & 0.50$\pm0.02$       & 1.47$\pm0.05$  &  0.99$\pm$0.03 & 0.43$\pm$0.02 \vspace{1mm}\\        
Si &    0.53$\pm$0.02  & 1.03$\pm0.09$  & 0.76$\pm0.03$   & 1.65$\pm0.07$ & 1.19$\pm$0.03 &  0.49$_{-0.03}^{+0.02}$ \vspace{1mm}\\        
S  &    0.42$\pm$0.02 & 0.28$\pm0.12$  & 0.80$\pm0.05$       & 1.03$\pm0.06$ & 0.69$\pm$0.04 & 0.64$_{-0.07}^{+0.06}$ \vspace{1mm}\\ 
Fe &    0.33$\pm$0.03  & 0.49$_{-0.02}^{+0.09}$  & 0.33$\pm0.01$      & 0.32$\pm0.07$ & 0.39$\pm$0.02 & 0.33$\pm$0.01 \vspace{1mm} \\
\hline
 $\chi^{2}$    & 11081.6 &    & 10826.9  &  & 10955.9 & \\
  d.o.f. & 9340 &  & 9340 &  & 9340 &  \\
\noalign{\smallskip}\hline\noalign{\smallskip}
\end{tabular}
\end{center}
\label{tab:abundances_2}
\end{table*}

\begin{table*}
\caption[]{Elemental abundances of \src from \ero compared with EPIC-pn 2019 observation. The elemental abundances are tied between different shock components.}
\begin{center}
\begin{tabular}{lccccc}
\hline\hline\noalign{\smallskip}
\multicolumn{1}{l}{Element} &
\multicolumn{2}{c}{\ero} &
\multicolumn{1}{c}{\ero} &
\multicolumn{1}{c}{\ero + } &
\multicolumn{1}{c}{EPIC-pn 2019 } \\
   &                             Comm.                 &   FL                   &     Comm. + FL & EPIC-pn 2019 & \\
\noalign{\smallskip}\hline\noalign{\smallskip} 
N  & 2.13$_{-0.45}^{+0.53}$ & 4.79$_{-0.74}^{+0.88}$ & 3.72$_{-0.43}^{+0.49}$ & 3.68$_{-0.46}^{+0.54}$ & 3.40$_{-1.52}^{+3.40}$ \vspace{1mm} \\
O & 0.21$\pm0.04$ & 0.41$_{-0.05}^{+0.06}$ & 0.32$\pm0.04$ & 0.32$\pm0.03$ & 0.23$_{-0.10}^{+0.19}$ \vspace{1mm} \\        
Ne & 0.56$_{-0.09}^{+0.10}$ & 0.92$_{-0.10}^{+0.11}$ & 0.63$\pm0.07$ & 0.77$_{-0.07}^{+0.05}$ &  0.76$_{-0.32}^{+0.40}$ \vspace{1mm}  \\      
Mg & 0.58$\pm0.07$ & 0.76$_{-0.07}^{+0.03}$ & 0.69$_{-0.05}^{+0.04}$  & 0.69$_{-0.04}^{+0.05}$ & 0.68$_{-0.26}^{+0.27}$    \vspace{1mm}\\      
Si         & 0.85$\pm0.07$ & 0.98$_{-0.06}^{+0.03}$ & 0.94$\pm0.05$  & 0.93$_{-0.03}^{+0.02}$ &   0.90$_{-0.29}^{+0.25}$ \vspace{1mm}\\       
S    & 0.71$_{-0.09}^{+0.10}$ & 1.07$_{-0.08}^{+0.08}$ & 0.94$_{-0.06}^{+0.07}$ & 0.89$_{-0.05}^{+0.06}$ &  0.65$_{-0.21}^{+0.25}$ \vspace{1mm}\\       
Fe          & 0.42$_{-0.01}^{+0.01}$ & 0.38$_{-0.02}^{+0.02}$ & 0.39$\pm0.02$ &  0.31$\pm0.02$ & 0.31$_{-0.07}^{+0.15}$ \vspace{1mm}\\        
\noalign{\smallskip}\hline\noalign{\smallskip}
\end{tabular}
\end{center}
\label{tab:abundances_ero}
\end{table*}

\subsection{\ero spectra and simultaneous fitting of EPIC-pn 2019, \ero Comm, and \ero FL spectra}
\label{sec:simulfit}
\begin{figure*}
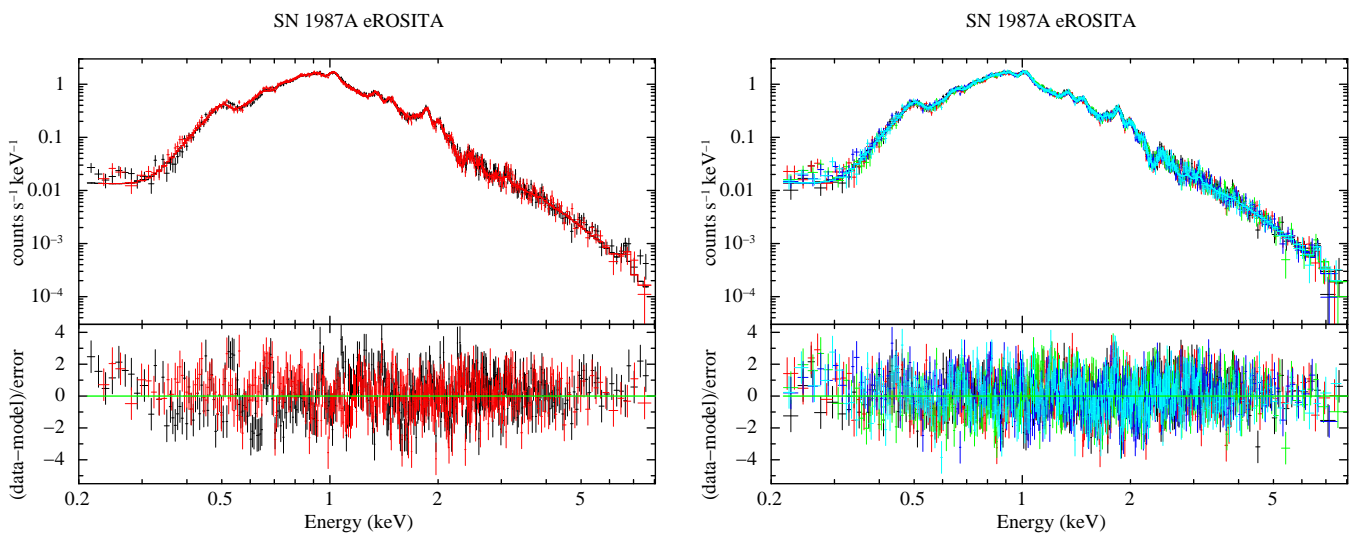

 \begin{center}
  \resizebox{0.49\hsize}{!}{\includegraphics[angle=-90]{model_3vpshock_dewey_tau0_PAT15_com_TM34_c001.ps}}
 \resizebox{0.49\hsize}{!}{\includegraphics[angle=-90]{model_3vpshock_dewey_tau0_PAT15_FL_TM12346_c001.ps}}
  \end{center}
  \caption{Left: \ero spectra (TM3, TM4) from the commissioning observation 700016 with the best-fit model obtained from the combined fit (\texttt{PAT} = 15).  Right: Same for the first-light observation 700161. The spectra from TM1, 2, 3, 4, and 6 were fit simultaneously.}
  \label{fig_ero_sim}
\end{figure*}

The \ero spectra were fitted with the model used for the EPIC-pn 2019 observation, with elemental abundances, normalisations, and the ionisation timescales as free parameters, but again linked between the different \ero spectra. As a first exercise, we fitted the \ero spectra from the two observations (\ero Comm and \ero FL)  separately. Only an inter-calibration constant was allowed to take a possible difference in the normalisations between the different TMs into account (which might be caused by problems in the derivation of good time intervals (GTIs) at this early stage of data analysis, e.g.). The best-fit spectral parameters for the two observations are given in Tables~~\ref{tabxayspec} and \ref{tab:abundances_ero}. The spectra from the two observations extracted using all valid-pattern events  (\texttt{PAT} = 15) are shown in Fig.~\ref{fig_ero_sim}. Next, we also fitted all the \ero spectra from the two observations simultaneously and verified  that the spectral fits resulted in reasonable values of spectral parameters and flux estimates between the different TMs.

\begin{figure*}
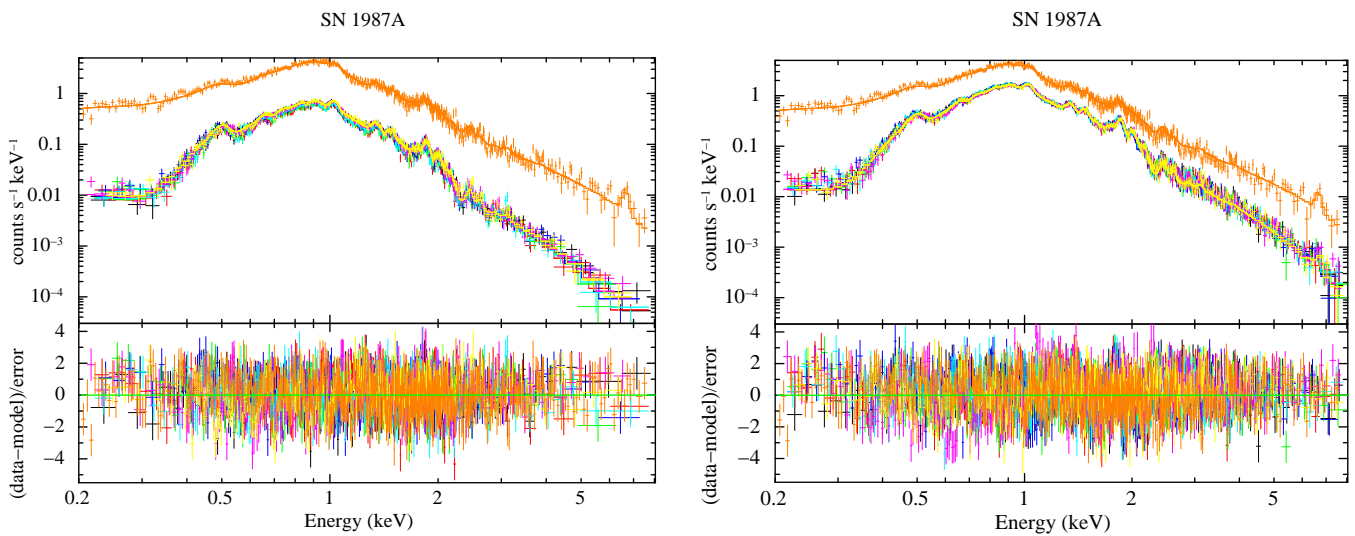

  \begin{center}
  \resizebox{0.49\hsize}{!}{\includegraphics[angle=-90]{model_3vpshock_dewey_tau0_PAT1_simul_TM_c001_xmm_2019_xmm_2019_err.ps}}
   \resizebox{0.49\hsize}{!}{\includegraphics[angle=-90]{model_3vpshock_dewey_tau0_PAT15_simul_TM_c001_xmm_2019_xmm_2019_err.ps}}
  \end{center}
  \caption{\ero spectra (left: single-pixel events, right: \texttt{PAT} = 15) from the combined fit of the commissioning observation 700016 and first-light observation 700161 and 2019 EPIC-pn (orange) spectrum fitted simultaneously.}
  \label{fig_ero_xmm_sim}
\end{figure*}

\begin{table*}
\caption{Comparison of observed fluxes between the different \ero telescope modules using all valid patterns with the best-fit spectral model (obtained by varying the best-fit spectrum for EPIC-pn).}
\begin{tabular}{lcccccc}
   \hline\hline\noalign{\smallskip}
   Observation & Energy band & \multicolumn{5}{c}{Flux (\oergcm{-13})} \\
               &  (keV)      & TM1 & TM2 & TM3 & TM4 & TM6 \\
   \hline\noalign{\smallskip}
   \ero Comm. & 0.5--2.0 &                &                & 68.6~$^{+1.6}_{-1.9}$ & 68.2~$^{+1.6}_{-1.9}$  & \\
          & 3.0--10.0 &                &                & 19.5~$^{+1.0}_{-0.8}$ & 19.4~$^{+1.1}_{-0.9}$  & \\
   \ero FL & 0.5--2.0 & 68.9$\pm$1.2 & 66.1$\pm$1.0 & 67.2$\pm$1.0 & 68.0$\pm$1.1 & 66.5$\pm$1.0 \\
          & 3.0--10.0 & 18.7$\pm$0.8 & 18.0$\pm$0.7 & 18.2$\pm$0.7 & 18.5$\pm$0.8 & 18.1$\pm$0.7 \\
   \hline
   \label{tabxray}
\end{tabular}
\tablefoot{
Fluxes derived from \texttt{PAT} = 15 spectra. 
Errors are quoted at 90\% confidence.
}
\label{taberoflux}
\end{table*}

The \ero spectra from the two observations were finally fitted jointly with the EPIC-pn spectrum from the 2019 observation, which was performed six weeks before the \ero FL observation. The spectra with best-fit model are shown in Fig.~\ref{fig_ero_xmm_sim}, and the best-fit parameters are also listed in Tables~\ref{tabxayspec} and \ref{tab:abundances_ero}. The fluxes in the 0.5-2\,keV and 3-10\,keV band for each \ero observation and TM are summarised in Table~\ref{taberoflux}. The difference in flux between the individual TMs reflects the inter-calibration constants between the different telescope modules in the two observations.
Fluxes obtained from the \ero data and the 2019 \xmm EPIC-pn data agree within 4\% and 7\% in the energy range of 0.5--2\,keV for all-valid and single-pixel events, respectively. The corresponding values in the energy range of 3--10\,keV are 10\% and 3\%. The difference in fluxes obtained between the different TMs is $\sim4\%$ and point to the level of systematics in the averaged flux value from \ero used to trace the flux evolution of \src in Sect.~\ref{sec:flux}.
The fluxes obtained from the \ero spectra using valid-pattern and single-pixel events disagree. This is caused by inconsistencies in pattern-fractions and is an \ero calibration issue that is currently (April 2021) under investigation. The model parameters predict a slightly lower post-shock temperature $kT_3$, and a slightly higher upper limit of the ionisation ages than inferred from the EPIC-pn 2019 observation alone. The elemental abundances of Ne, Si, and S are also slightly higher than derived from EPIC-pn spectra. The spectral parameters obtained by the joint fit of the \ero and EPIC-pn spectra broadly agree with the parameters obtained from the EPIC-pn 2019 spectrum, however.


We also tested the statistical significance of the third thermal component in the three-component shock model using the \ero data. We fitted the seven spectra obtained from the \ero Comm and FL observations with only two \texttt{VPSHOCK} components as in the case of \xmm EPIC-pn. The two-component fit results in a $\chi^{2}$ of 2417.6, to be compared with 2390.8 in case of the three-component model for 1941 degrees of freedom. The addition of a power-law component instead of a hot plasma component to the two-component shock model does not lead to any improvement in $\chi^{2}$ (2417.6 for 1935 degrees of freedom).
This further supports our claim that the three-component plane-parallel shock model is favoured for the spectral modelling of \src.

\subsection{Flux evolution of \src}
\label{sec:flux}
\begin{figure*}
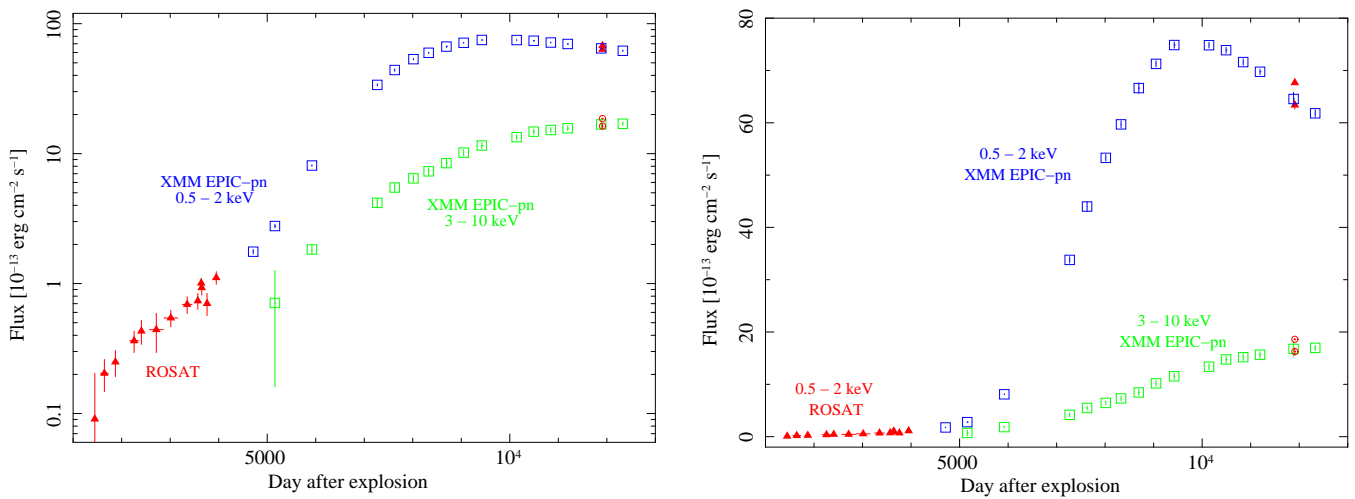

  \begin{center}
  \resizebox{0.49\hsize}{!}{\includegraphics[angle=-90]{SN1987A_lc_data_soft_hard_20201124_XMMonly.ps}}
  \resizebox{0.49\hsize}{!}{\includegraphics[angle=-90]{SN1987A_lc_data_soft_hard_20201124_XMMonly_lin.ps}}
  \end{center}
  \caption{X-ray light curve of \src in logarithmic (left) and linear scale (right). \ero fluxes are the error-weighted average flux derived from the simultaneous fit to the spectra of the commissioning and the first-light observation. Fluxes derived from single-pixel events are somewhat below the EPIC-pn values, while PAT15 events give higher fluxes.}
  \label{fig1987Alc}
\end{figure*}

In order to trace the flux evolution of \src, the three-temperature shock model described in Sect.~\ref{sec:xmmspec} was used because it provides an acceptable fit to the data.
With the observed fluxes obtained from the spectral fits, the long-term X-ray flux evolution of \src in the soft (0.5--2\,keV) and hard (3--10\,keV) X-ray bands is shown in Fig.~\ref{fig1987Alc}, including the archival results from ROSAT reported by \cite{1994A&A...281L..45B} and \cite{1996A&A...312L...9H}. 
A flattening and then turnover of the soft X-ray flux is observed between days 10142–10493, followed by a steady decline. However, the hard
X-ray flux continues to increase, but at a slower rate. A comparison between the flux obtained with 2019 EPIC-pn and the \ero observations is also marked in the figure. 
The error-weighted average flux obtained using only single-pixel patterns (\texttt{PAT} = 1) from \ero is systematically lower than obtained with EPIC-pn. On the other hand, the corresponding flux obtained using 
all valid patterns (\texttt{PAT} = 15) from \ero is slightly higher than obtained with EPIC-pn. Because the lower fraction of single-pixel events is largely compensated for by a higher fraction of double-pixel events, the \ero fluxes derived from all valid patterns are likely more reliable. See Tables~\ref{tabobsxmm} and ~\ref{taberoflux} for the flux values.

\section{Discussion}
\label{sec:duscussion}

This paper reports the most complete and most recent flux  evolution of \src in X-rays starting from 1991 using {\it ROSAT} (1448\,d after the explosion, when the source became first visible in soft X-rays) up to November 2020 (12329\,d after explosion) using \xmm and \ero observations during its commissioning phase and first light.  The \xmm observations are  part of a monitoring campaign since 2007 and were presented up to 2012 in \citet{2006A&A...460..811H}, \citet{2008ApJ...676..361H}, \citet{2010A&A...515A...5S}, and \citet{2012A&A...548L...3M,2016A&A...585A.162M}.  An additional \xmm observation to allow cross-calibration studies between \xmm and \ero was performed in September 2019. Another observation from November 2019 (not used in this work; see Sect.~\ref{sec:observations}) was presented in \cite{2021arXiv210302612A} and \cite{2021arXiv210303844S}. We  used the EPIC-pn monitoring observations to trace a possible temporal evolution of the different shocked plasma components and their elemental abundances over the past 14 years. Our analysis demonstrates that a three-temperature (cooler, warm, and hot) shock model is statistically favoured over a two-temperature model, in line with the findings first reported in \cite{2010MNRAS.407.1157Z}, \citet{2012ApJ...752..103D} and thereafter used in  \citet{2012A&A...548L...3M} and recently in \cite{2021arXiv210302612A}, with slightly different temperatures for the shock components, especially the cooler component. No evidence of an additional non-thermal component is found in the spectrum of \src, consistent with the analysis of the joint \xmm and NuSTAR observations by \cite{2021arXiv210302612A}.

\subsection{Clear trend of continued flux decline in the soft X-ray band and comparison of fluxes between \xmm and \ero}
The X-ray light curve of \src in the 0.5--2\,keV band flattened around 9424\,d (2012 EPIC-pn) and then displayed a turnover between 10142--10493\,d (2014--2015 EPIC-pn). Around the same time, a break is visible in the slope of the increasing hard-band flux (3--10\,keV). Thereafter, while the soft-band flux has shown a continuous decline up to the last \xmm monitoring observation in November 2020, the hard-band flux has continued to increase, but with a flatter slope. The described trend in the fluxes between the soft and the hard band implies that the blast wave has now passed beyond the dense structures of the equatorial ring and is expanding farther into a more tenuous CSM. The measured soft-band flux from 12329\,d (Nov. 2020) indicates that the flux has dropped to $\sim$6.2\ergcm{-12} , which is 18\% below the maximum value measured on  9424\,d (2012). The recent flux evolution confirms that the flux levels off in the soft-band light curve around 9000\,d, as reported in \cite{2016ApJ...829...40F}, and clearly demonstrates the continued flux decline.

An additional aim of the 2019 EPIC-pn observation was to allow cross-calibration studies between \xmm and the early \ero observations of \src performed in September and October 2019. The \ero fluxes (which are the error-weighted average flux derived from the simultaneous fit to the spectra of
the commissioning and the first-light observation) agree with the 2019 EPIC-pn fluxes within 4\% and 7\% in the energy range of 0.5--2\,keV for all-valid and single-pixel events, respectively. The corresponding values are 10\% and 3\% in the energy range of 3--10\,keV. The discrepancy between the \ero fluxes obtained using the valid-pattern and single-pixel events is caused by inconsistencies in pattern fractions, as discussed in Section~\ref{sec:simulfit}.

\subsection{Origin and temporal evolution of the plasma components in \src }
The physical origin of the different plasma components in \src is complicated due to the complex CSM structure and the shock interactions. The likely origin of the thermal X-ray emission includes the dense clumps of the equatorial ring, the lower-density inter-clump gas (also including the destroyed or evaporated clumps), an ionised HII region surrounding the ring, as well as reflected shock-heated gas from the equatorial ring region and the reverse shock moving into the ejecta. Previous studies that were based on X-ray data at the peak of the interaction of the blast wave with the dense equatorial ring attributed the low-temperature plasma to the interaction of the blast wave with the dense equatorial ring and the higher-temperature components to the emission from the inter-clump gas, from the surrounding HII regions as well as from the reflected shocks \citep{2006ApJ...646.1001P,2010MNRAS.407.1157Z,2012ApJ...752..103D}. However, in the recent epochs that are included in our analysis, the emission component from the reverse shock moving into the ejecta is expected to have a growing contribution to the hard X-ray emission modelled by the hot plasma component. This is corroborated by recent hydrodynamical simulations, which indicate that the X-ray emission from \src will start to be dominated by the SN ejecta heated by the reverse shock in 32--34 years after the explosion \citep{2015ApJ...810..168O,2020ApJ...899...21B}.
In this work, we attribute the cooler, warm, and hot components of the three-temperature \texttt{VPSHOCK} model to trace the emission caused by the blast-wave shock transmitted into dense gas of the circumstellar ring including the inter-clump gas, the blast wave propagating in the low-density CSM/HII region, and the reverse shock evolving into the ejecta, respectively. Our motivation for the origin of the plasma components is mainly driven by the evidence of a different  chemical composition of the warm and hot plasma components, especially O, Ne, Mg, and Si. This is discussed further in Sect.~\ref{dis:abun}. 

The temporal evolution of the spectral parameters of the three plasma components in \src is plotted in Figs.~\ref{fig1987Atemp} and \ref{fig1987Anorms}. A gradual increase in the electron temperature of the cooler component is observed up to 11193\,d. This has been reported previously for the epochs 2003--2007 using \cxo observations \citep{2006ApJ...646.1001P} and \xmm RGS observations \citep{2010A&A...515A...5S}. The increasing trend can be explained if regions with slightly higher temperatures contributed more to the emission measure at later epochs,  for instance if the emission were dominated by the shocked dense clumps at the beginning, but was increasingly overtaken by the lower-density gas inside the circumstellar ring with time, which might constitute both the inter-clump gas and the evaporated clump material. No clear trend in the electron temperature of the hot component can be inferred.

 A gradual increase in shock ionisation age is seen, as expected if the blast wave encounters gas of increasing density as it propagates into the circumstellar ring structure. It seems to increase also after 11193\,d, most likely because of the evaporation of clumps in the ring as well as the evolution of the reverse shock. A linear fit to the upper limit of the ionisation timescale from 7269--12329\,d leads to an average density of 3577$\pm143$ cm$^{-3}$ for the cooler plasma component. This is consistent with what was found by \citet{2009PASJ...61..895S} based in the first three \xmm observations in Table~\ref{table_EPICobs}. This value is also in line with the estimates obtained from early optical analysis of the ionised gas, which predicted a density of $1-3\times10^{4}$ cm$^{-3}$ for the clumps and a density of $1-6\times10^{3}$ cm$^{-3}$ for the inter-clump gas \citep{1996ApJ...464..924L} and from hydrodynamical simulations \citep{2015ApJ...810..168O}. The obtained density range further suggests that the cooler plasma component is dominated by the inter-clump gas.  However, we would like to add a note of caution that the estimate of the density is derived from the \texttt{vpshock} model, which assumes an adiabatic, one-dimensional plasma shock propagating into a uniform CSM, which is an oversimplified picture in the case of the dense equatorial ring.
 
The normalisations norm$_1$, norm$_2$, and norm$_3$ of the cooler, warm, and hot components of the \texttt{VPSHOCK} model trace the emission caused by the blast-wave shock transmitted into the dense equatorial ring gas, the blast wave propagating in the CSM, and the reverse shock evolving into the ejecta, respectively. The contribution of norm$_2$ increases with time, but shows indications of flattening after 11193\,d. Comparison with the hydrodynamical simulation and synthetic light curves to predict the SNR evolution from  \cite{2020A&A...636A..22O} shows that the observations match the scenario of a remnant evolution for a blue supergiant progenitor well. The increasing contribution of norm$_3$ in the recent years, while norm$_2$ stays constant and norm$_1$  decreases may reflect the fact that the emission caused by the forward shock is leaving the equatorial ring and the reverse shock in the ejecta is becoming more dominant. This is in support of the origin of the hot plasma component as the reverse-shock emission component.

We verified that the time evolution of the parameters is not affected by the assumption of fixing $kT_{2}=1.15$\,keV in our spectral model. In order to test this, we repeated the exercise with fixing  $kT_{2}=1.0$\,keV and $kT_{2}=1.3$\,keV. A marginal increase in both $kT_{1}$ and $kT_{3}$ and a corresponding decrease in  $\tau_{\rm u,}$ is seen with increasing $kT_{2}$ from 1.0 to 1.1 to 1.3\,keV are really all temperatures increasing when $kT_{2}$ is forced to increase?). No change in the normalisation values are detected within the errors. More importantly, the pattern in the evolution of the parameters remains the same in all cases, establishing that the time evolution does not depend on the exact choice of  $kT_{2}$. This has also been confirmed previously by \cite{2012ApJ...752..103D}, who found that the exact choice of the middle temperature is somewhat arbitrary and degenerate with other parameters. We further caution that the ionisation ages are tied to each other to cover a factor of 2, and this could lead to some systematics in some parameter estimates, especially the density of the plasma components.

\subsection{Elemental abundances}
\label{dis:abun}
The elemental abundances of the different plasma components were studied by setting the abundances of the warm and hot plasma component free with respect to the cooler component. The simultaneous fit from EPIC-pn favours a higher abundance for the warm and hot components (especially the hot component and for the elements O, Ne, Mg, and Si). A time evolution of the elemental abundance was also explored by dividing the data into two groups: before the X-ray emission was dominated by the shocked plasma from the dense equatorial ring (2007--2015), and after (2016--2020). The data show evidence of a lower abundance of N at the later epoch, together with a marginal increase in the abundance of Ne and Mg. No other clear trend can be seen. However, we note that the elemental abundances determined in epoch 2016--2020 are time-averaged over the three plasma components and the contribution of the hot plasma (which is indicated to have the highest abundances) became dominant in the observation of the latest epoch 12329\,d, as seen in  Fig.~\ref{fig1987Anorms}.
\citet{2020ApJ...899...21B} obtained elemental abundances of Ne = 0.4, Mg = 0.34, Si = 0.4, S = 0.34, and Fe = 0.16 from \cxo HETG/LEG. The latest \xmm study by \cite{2021arXiv210303844S} yields slightly higher abundances and also indicates that the elemental abundances have decreased in the latest epochs. While the analysis of \citet{2020ApJ...899...21B} was limited to softer energies below 3.0 keV, \cite{2021arXiv210303844S} analysed the RGS and EPIC data together, but assumed a model consisting of two thermal emission components. Therefore these studies miss the emission from the hottest component, which is likely caused by the reverse shock propagating into the ejecta. This is the component for which we therefore expect higher element abundances.  Future analysis including the grating spectra will be important to confirm these results.

Our analysis of the \xmm EPIC and \ero spectra by decoupling the element abundances of the warm or hot component (Table \ref{tab:abundances_2}) shows that the abundances for the cool and warm component are consistent with those obtained from the grating spectra when the abundances in the hot component are left free. 
Moreover, the abundances in the hot component are significantly higher in that case, supporting the ejecta scenario. Iron is the exception, with a low abundance. This can be expected if the reverse shock has not yet reached and heated the inner ejecta, which contains the outer layer of iron that escaped fall-back on the newly born neutron star. In this scenario, we predict that the abundance of Fe (and other heavier elements such as Si and S) in the hot component will rise in future observations.
\section{Conclusions}
\label{sec:conclusions}
\src has shown dramatic evolution in the X-ray spectrum and flux over the past $\sim$15 years. The soft-band flux after 10142--10493\,d shows a continued decline and has decreased by 18\% of its peak value in November 2020. The X-ray emission caused by the forward shock after traversing the equatorial ring and by the reverse shock in the ejecta is becoming more dominant, and the SNR has entered a new evolutionary phase.
During its ongoing all-sky surveys, SRG/\ero will continue to monitor \src. This will provide crucial information about the new evolutionary phase of the supernova remnant. Changes in the emission from the forward and reverse shocks as well as in abundances and N$_{\rm H}$ are expected and will be followed with SRG/\ero in the next  years. This will be complemented by our continued monitoring of the source with EPIC-pn.


\begin{acknowledgements}

We thank the referee for a very thorough and critical reviewing of the manuscript which helped in considerably improving the clarity of the paper.

This work is based on data from \ero, the primary instrument aboard SRG, a joint Russian-German science mission supported by the Russian Space Agency (Roskosmos), in the interests of the Russian Academy of Sciences represented by its Space Research Institute (IKI), and the Deutsches Zentrum f{\"u}r Luft- und Raumfahrt (DLR). The SRG spacecraft was built by Lavochkin Association (NPOL) and its subcontractors, and is operated by NPOL with support from the Max Planck Institute for Extraterrestrial Physics (MPE).
The development and construction of the \ero X-ray instrument was led by MPE, with contributions from the Dr. Karl Remeis Observatory Bamberg \& ECAP (FAU Erlangen-N{\"u}rnberg), the University of Hamburg Observatory, the Leibniz Institute for Astrophysics Potsdam (AIP), and the Institute for Astronomy and Astrophysics of the University of T{\"u}bingen, with the support of DLR and the Max Planck Society. The Argelander Institute for Astronomy of the University of Bonn and the Ludwig Maximilians Universit{\"a}t Munich also participated in the science preparation for \ero.
The \ero data shown here were processed using the eSASS/NRTA software system developed by the German \ero consortium.
The \ero data shown here were processed using the eSASS software system developed by the German \ero consortium.  M.S. acknowledges support by the Deutsche Forschungs-gemeinschaft through the Heisenberg professor grant SA 2131/12-1.
\end{acknowledgements}

\bibliographystyle{aa} 
\bibliography{general} 

\end{document}